\def \SAIT #1 #2 {{\em Mem.\ Soc.\ Astron.\ It.\/} {\bf #1}, #2}
\def \MESS #1 #2 {{\em The Messenger\/} {\bf #1}, #2}
\def \ASTRNACH #1 #2 {{\em Astron. Nach.\/} {\bf #1}, #2}
\def \AAP #1 #2 {{\em Astron. Astrophys.\/} {\bf #1}, #2}
\def \AAL #1 #2 {{\em Astron. Astrophys. Lett.\/} {\bf #1}, L#2}
\def \AAR #1 #2 {{\em Astron. Astrophys. Rev.\/} {\bf #1}, #2}
\def \AAS #1 #2 {{\em Astron. Astrophys. Suppl. Ser.\/} {\bf #1}, #2}
\def \AJ #1 #2 {{\em Astron. J.\/} {\bf #1}, #2}
\def \ANNREV #1 #2 {{\em Ann. Rev. Astron. Astrophys.\/} {\bf #1}, #2}
\def \APJ #1 #2 {{\em Astrophys. J.\/} {\bf #1}, #2}
\def \APJL #1 #2 {{\em Astrophys. J. Lett.\/} {\bf #1}, L#2}
\def \APJS #1 #2 {{\em Astrophys. J. Suppl.\/} {\bf #1}, #2}
\def \APSS #1 #2 {{\em Astrophys. Space Sci.\/} {\bf #1}, #2}
\def \ASR #1 #2 {{\em Adv. Space Res.\/} {\bf #1}, #2}
\def \BAIC #1 #2 {{\em Bull. Astron. Inst. Czechosl.\/} {\bf #1}, #2}
\def \JSQRT #1 #2 {{\em J. Quant. Spectrosc. Radiat. Transfer\/} {\bf #1}, #2}
\def \MN #1 #2 {{\em Mon. Not. R. Astr. Soc.\/} {\bf #1}, #2}
\def \MEM #1 #2 {{\em Mem. R. Astr. Soc.\/} {\bf #1}, #2}
\def \PLR #1 #2 {{\em Phys. Lett. Rev.\/} {\bf #1}, #2}
\def \PASJ #1 #2 {{\em Publ. Astron. Soc. Japan\/} {\bf #1}, #2}
\def \PASP #1 #2 {{\em Publ. Astr. Soc. Pacific\/} {\bf #1}, #2}
\def \NAT #1 #2 {{\em Nature\/} {\bf #1}, #2}

\documentstyle{memsait}
\input epsf.sty
\begin{opening}
\title{EVOLUTION OF AGN AND STELLAR FORMATION THROUGH RADIO SURVEYS}
\author{Loretta Gregorini$^{1,2}$, Isabella Prandoni$^{1}$}
\institute{$^1$Istituto di Radioastronomia CNR, Bologna, Italy\\
$^2$ Dipartimento di Fisica, Universit\'a di Bologna, Italy}
\date{} 
\end{opening}

\begin{document}

\oddpagefooter{}{}{} 
\evenpagefooter{}{}{} 
\ 
\bigskip

\begin{abstract}
The change in the slope of the radio source counts
suggests the emergence of a new population of radio galaxies at mJy and 
sub-mJy levels. 
Our understanding of such faint radio sources has advanced over the last 
decade through increasingly sensitive radio surveys and follow-up works at 
optical wavelength. 
The sub--mJy population seems to include both star forming galaxies and 
classical (AGN-powered) radio sources, but the relative importance of the 
two classes is still debated. Recent results are reviewed and discussed.
\end{abstract}

\section{Introduction}
Differential radio source counts derived from deep 1.4 GHz surveys 
show a flattening below a few milliJanskys (Fig. 1). 
This change in slope is usually 
interpreted as the result of the emergence of a new population  
which does not appear at higher flux densities, where the counts are believed
to be dominated by the classical powerful radio galaxies 
and quasars, triggered by an active galactic nucleus (AGN).

To explain the new faint radio population several scenarios have been invoked:
strongly-evolving normal spirals ({\it e.g.} Condon 1984, 1989); a 
non-evolving population of local (z $<$ 0.1) low-luminosity galaxies 
({\it e.g.} Wall et al. 1986);
actively star forming galaxies ({\it e.g.} Rowan-Robinson et al 1993). 
The latter scenario is
strongly supported by extended optical identification works, 
showing that 
the counterparts of faint radio sources are often blue, disk galaxies 
({\it e.g.} Thuan and Condon 1987), with disturbed morphology and/or in 
merging galaxy systems.
The energy source in such galaxies is ultimately stellar, possibly entirely 
from supernovae explosions, widespread over the whole galaxy. 

However, nuclear emission could be important also at low radio fluxes. 
Both low luminosity AGNs and nuclear star formation have been suggested 
as possible mechanisms for producing the radio emission. 
In fact, to determine the dominant source
in any galaxy it is necessary to obtain from observations its radio morphology
and brightness, together with the morphology and spectrum of the 
optical counterpart.


\begin{figure}
\epsfysize=12cm 
\hspace{3.5truecm}\epsfbox{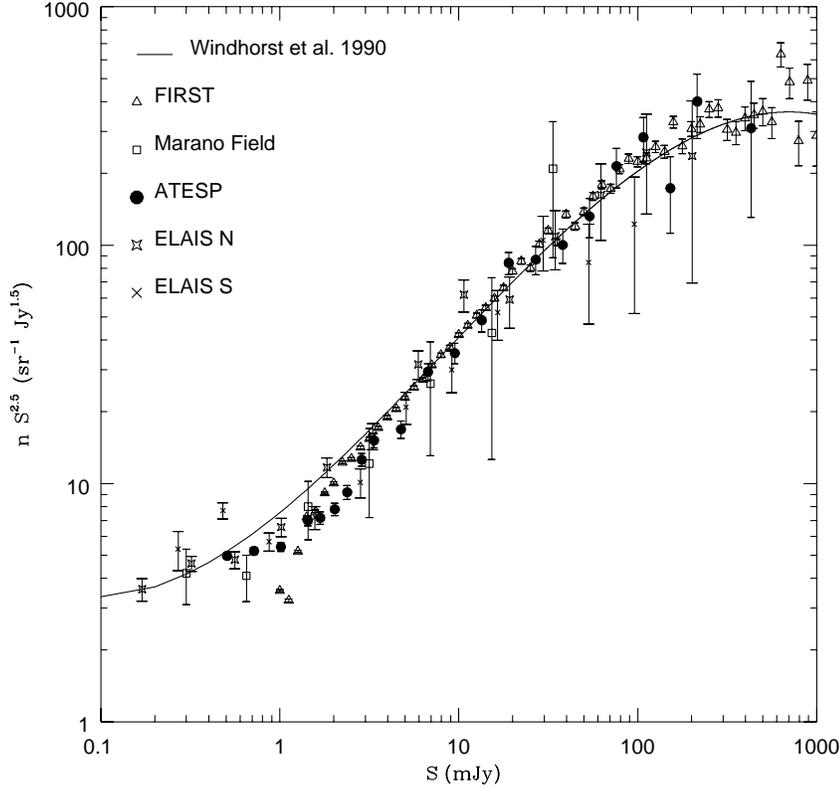} 
\caption[h]{Radio source counts at 1.4 GHz (normalized to Euclidean ones) as 
       a function of flux for 
       different samples (see Tab.~1): FIRST (triangles),  
       Marano Field (squares), ATESP (filled 
       circles), ELAIS North (stars) and ELAIS South (crosses).  
       The fit obtained by Windhorst et al. (1990) is also shown
       (solid line)}
\end{figure}

\section{Active Galactic Nuclei}

Radio and optical quasars, Seyfert galaxies, radio galaxies and blazars 
constitute a variety of astrophysical phenomena which may be interpreted 
as due to the existence of a massive black-hole (BH) in galaxy nuclei.
Powerful spectrographs and high-resolution imagers on optical (HST) and
X-ray telescopes (ROSAT and ASCA) allow to test the presence of a BH 
($M_{BH} \sim 10^6 - 3 \times 10^9 M_{sun}$) in a few nearby galaxies,
via gas and stellar dynamics. Models on quasars ({\it e.g.} Cavaliere and Padovani 
1989) suggest that the AGN phenomenon is short-lived and probably recurrent,
implying that a significant fraction of massive galaxies should contain a
inactive BH. 
Recently Vercellone and Franceschini (1999) explored the dependence 
of the galaxy 
nucleus emissivity at various wavelengths on the BH mass, using estimates of 
$M_{BH}$ for a sample of nearby galaxies. 

\begin{minipage}[t]{12.5cm}

\centerline{\bf Tab. 1}
\centerline{\bf Faint 1.4 GHz Radio Surveys}
\hspace{1.5cm} 
\begin{tabular}{|ll|ll|}
\hline
 Survey  & References &  Survey  & References\\             
\hline

NVSS & Condon et al. 1998  & PDF & Hopkins et al. 1998\\
FIRST & White et al. 1997  & & Georgakakis et al. 1999 \\
VLA-NEP & Kollgaard et al. 1994 & B93\footnote{Sample of radio sources 
studied by Benn et al. (1993). It collects
sources from three deep 1.4 GHz radio fields: 0846+45 (Lynx 3A, Oort 1987), 
0852+17 (Condon \& Mitchell 1984) and 1300+30 (Mitchell \& Condon 1985).}
  & Benn et al. 1993  \\
ATESP & Prandoni et al. 1999 & Marano Field & Gruppioni et al. 1997\\ 
LBDS & Windhorst et al. 1984 & & Gruppioni et al. 1999a \\
 & Kron et al. 1985 & Lockman Hole & de Ruiter et al. 1998 \\
ELAIS N & Ciliegi et al. 1999 & 1300+30 & Mitchell \& Condon 
1985 \\
ELAIS S & Gruppioni et al. 1999b & HDF+HFF\footnote{Hubble Deep Fields + 
Flanking Fields region.} & Richards 1999  \\
\hline
\end{tabular}
\end{minipage}
\vspace{1cm} 

They found a tight relationship of 
the BH mass with both the nuclear and the total radio flux at 5 GHz, which 
is thus a very 
good tracer of a super-massive BH and a good estimator of its mass.

The fraction of sources below 1 mJy associated with elliptical galaxies
at $z \sim 1$ found in very deep radio surveys ({\it e.g.} Gruppioni et al. 
1997; Hammer et al. 1995) may be more distant examples and may offer the possibility 
to explore the 
evolution of the mass and distribution of black holes with cosmic epoch.

In a recent paper Ho (1999) used optical spectroscopic information for a 
sample of nearby early-type galaxies surveyed with the VLA, to establish 
the physical nature of the low-power radio cores present in these objects.
Comparison of the observed radio continuum power with that expected from the
thermal gas traced by the optical emission lines implies that the bulk of
radio emission is nonthermal. The relation between radio power and line 
emission observed in this sample is consistent with the low-luminosity 
extension of similar relations seen in classical radio galaxies and luminous 
Seyfert nuclei.  A plausible interpretation of this result is that the weak 
nuclear sources seen in these galaxies are simply the low-luminosity 
counterparts of more distant, luminous AGNs. 

\section{Star Forming Galaxies}

The prototype of nearby starburst galaxies is M82 (Muxlow et al 1994) which
shows the presence of many compact bright radio sources. Some of
these sources show large flux changes over scales of a few years suggesting 
that they are radio supernovae (SN) associated with the initial supernova 
explosion; while size measurements (typically 2-3 pc) indicate that the bulk 
of them are associated with young SN remnants. 
Typical star formation rates for the sub--mJy/mJy starbursts range from 1 to 
100 $M_{sun}~yr^{-1}$ (for massive stars with $M > 5 M_{sun}$). These star
formation rates can only be maintained for short periods ($< 10^{8-9} yr$)
before depleting the available gas in these systems. Thus, radio emission
which is almost extinction-free is a sensitive measure of recent starburst 
activity, where both UV and $H_{\alpha}$ observations are heavily affected by 
internal absorption. 
Its utility relies on the hypothesis that the radio 
luminosity is directly proportional to the supernova rate. The radio-FIR
correlation can be cited as support for this hypothesis. 
It is noteworthy that FIR samples are also extinction--free, but have very
low spatial resolution, making very difficult the identification 
follow--up. Moreover, radio samples allow to study star--forming galaxies 
at a flux density level considerably deeper than the 60 $\mu$Jy completeness 
limit of the IRAS Faint Galaxy Catalog and barely reachable in deep surveys
carried out by the ISO satellites ($S_{60\mu m} \sim 10--20$ mJy).

\begin{figure}[t]
\epsfysize=12cm 
\hspace{3.5truecm}\epsfbox{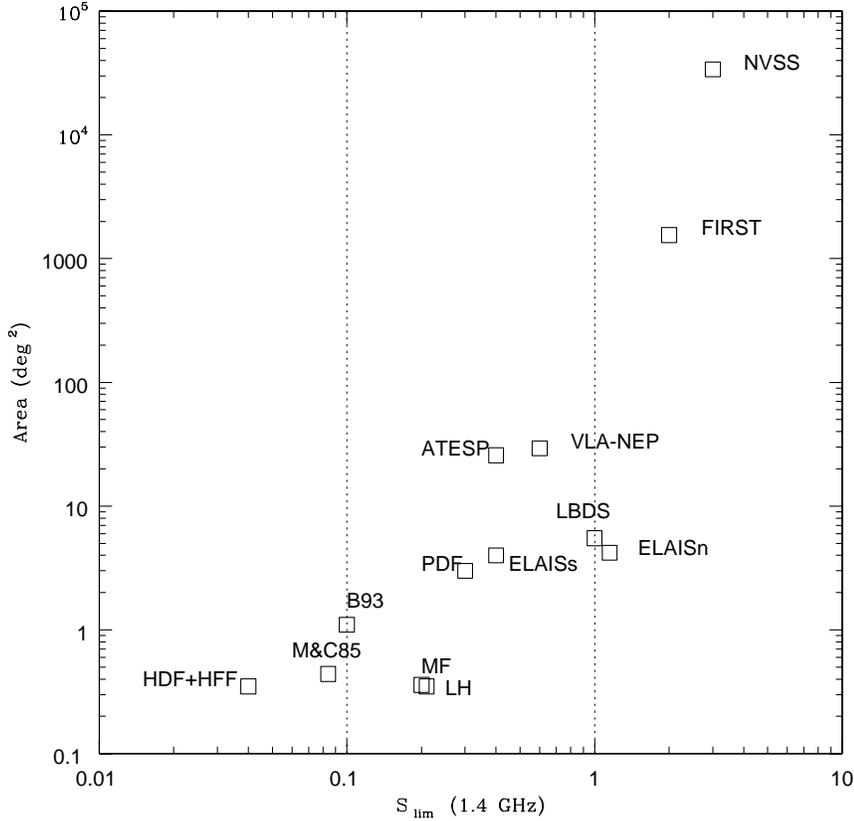} 
\caption[h]{Largest deep 1.4 GHz radio surveys: area covered (square degrees) 
vs. (80\% completeness) flux limit (mJy). The two 
vertical lines show the 1 mJy and 0.1 mJy flux limits.  
Some of the surveys reported here do reach deeper fluxes on sub-areas.
For simplicity, this extra-information is not plotted. For more information on
the surveys see references reported in Tab.~1.}
\end{figure}

\begin{figure}[t]
\epsfysize=12cm 
\hspace{3.5truecm}\epsfbox{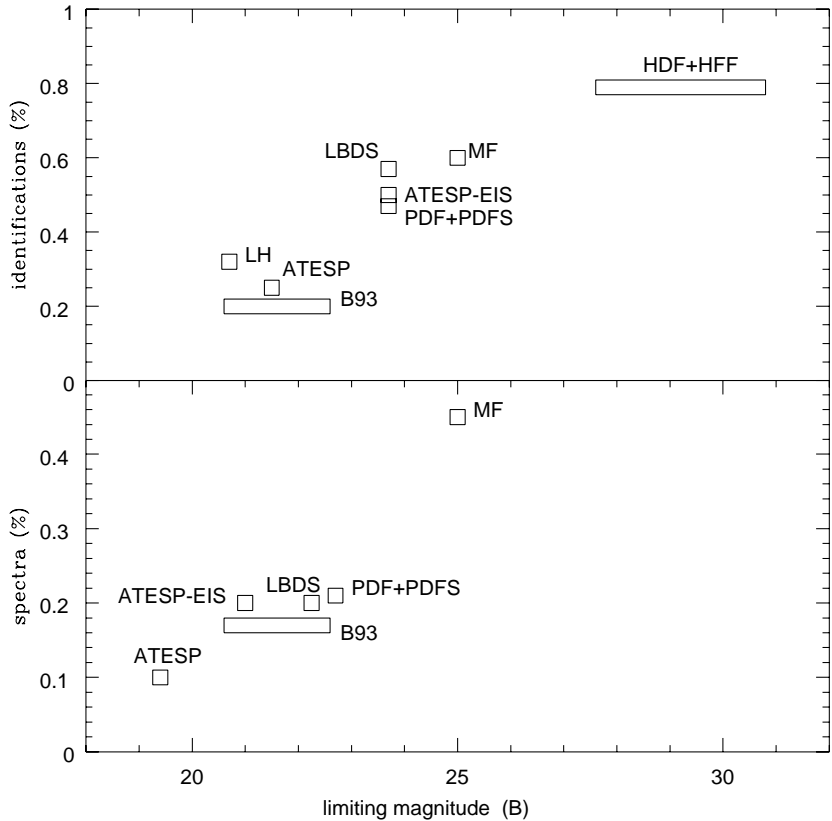} 
\caption[h]{1.4 GHz sub-mJy surveys: optical follow-up. Top panel: fraction of 
radio sources identified (i.e. optical counterpart found from imaging)
as a function of the limiting magnitude of the images. 
Bottom panel: fraction of radio sources with spectral information 
(redshift and possibly spectral classification) vs limiting magnitude
of spectroscopy. References are reported in Tab.~1. }
\end{figure}

\section{Faint Radio Surveys and Optical Follow--up}

Our understanding of the faint radio sources has advanced over the last 
decade through increasingly sensitive radio surveys and follow--up
works at optical wavelengths. The largest faint 1.4 GHz radio surveys 
available up to now are summarized in Tab.~1 and in Fig.~2, where  
we plotted the sky coverage versus the flux limit of the survey. 

The realization of a survey is time-consuming; therefore
the deeper the sensitivity, the smaller the area covered by the survey. For 
instance the deepest 1.4 GHz survey available (HDF+HFF in Fig.~2) is sensitive 
to $\mu$Jy sources (flux limit $\sim 40$ $\mu$Jy), but 
it covers only a fraction of square degree (0.35 sq. degr.). 

The ATESP survey (Prandoni et al. 1999), on the other hand, is a good 
compromise between flux 
limit ($\sim 0.4$ mJy) and area covered (about 26 square degrees). 
This radio survey overlaps the region of the ESO Slice Project (ESP) galaxy 
redshift survey (Vettolani et al. 1997). The catalogue contains about 
3000 radio sources, more than 1000 being sub-mJy sources. 

Despite the effort devoted to study the faint radio population there are still
open questions: (i) what is the stellar population in the sub--mJy/mJy
radio sources; (ii) which is the relation between star--formation 
activity and radio properties; (iii) which is the relation to the local 
population of normal galaxies. In order to address these points 
optical/near-infrared photometry and optical spectroscopy are needed. 

In Fig.~3 we summarize the optical follow-up available for the sub--mJy radio 
surveys shown in Fig.~2 (for references see Tab.~1). 
It is clear that for large radio samples it is very difficult to obtain 
deep optical imaging and spectroscopy and the fraction of sources identified  
is small. For instance $\sim 25\%$ of the ATESP radio sources have been 
identified, and only for $\sim 10\%$ spectra have been obtained. 
Nevertheless deeper imaging and spectroscopy is available for a 
sub-sample covering $\sim 3$ sq. degr. (ATESP-EIS, see Nonino et al. 1999).
Typically, no more than $\sim 50-60\%$ of  the radio sources in sub--mJy 
samples have been identified, even though in the Hubble Deep Field 
(HDF+HFF) an identification rate of about 80\% is reached.  
On the other hand the typical fraction of spectra available is only 
$\sim 20\%$. The best studied sample is the Marano 
Field, where $\sim 45\%$ of the sources have spectral information. 


\begin{table}[t]
\hspace{1.5cm} 
\centerline{\bf Tab. 2 }
\centerline{\bf mJy Population }
\begin{center}
\begin{tabular}{|l|c|c|c|c|c|}
\hline
  survey              &early-type &  late(S) &  AGN & others & $N_{tot}$ \\
                & \%       & \%  & \% & \% & \\
\hline
B93             & $45 \pm 20$ &  $27 \pm 16$ & $9 \pm 9$ & $18 \pm 13$ & 11 \\
PDF             & $62 \pm 14$ &  $17 \pm 7$ & $14 \pm 8$ & $7 \pm 5$ &  29 \\
ATESP-EIS       & $47 \pm 11$ & $8 \pm 5$ & $22 \pm 8$ & $24 \pm 8$ & 37 \\
\hline
\end{tabular}
\end{center}
\end{table}

\section{Results}

When discussing the results about the nature and physical properties of the
mJy and sub--mJy population, the numbers reported above must be kept in mind.
All the conclusions reached on the faint radio sources
are based on samples, for which photometry and spectroscopy are only available 
for a fraction of the whole population and selection effects can introduce 
serious biases.

It is therefore very important to compare homogeneous samples, i.e. 
with same radio flux limit and 
same limiting magnitude for the optical counterparts.
The role played by selection effects is clear when comparing
the results obtained in the Marano Field and the ones reached by Benn et al. 
(1993).
Gruppioni et al. (1999a) identified 44\% of all the radio sources fainter 
than 1 mJy with early-type galaxies; on the contrary, Benn 
et al. (1993) found a dominance of blue narrow emission line objects, 
identified as star--forming galaxies, and
a percentage of early-type galaxies of only about 8\%. This apparent 
discrepancy is probably due to the deeper optical magnitude reached in the 
identification work by Gruppioni et al.: the fraction of sub-mJy early-type 
galaxies in the Marano Field increases around $B = 22.5$, which is 
approximately the magnitude limit reached by Benn et al.

In order to compare the results obtained from different works we 
classified the faint radio galaxy population in three main 
classes:
(a) {\it early-type} galaxies (ellipticals and S0); (b) {\it late(S)} 
(star--forming galaxies); (c) {\it AGN} (objects with nuclear energy
source, i.e. Seyfert 2, Seyfert 1 and QSO); (d) {\it others} (objects which
cannot be classified in one of the three previous classes).

\begin{table}[t]
\centerline{\bf Tab. 3} 
\centerline{\bf sub--mJy Population }
\hspace{1.5cm} 
\begin{center}
\begin{tabular}{|l|c|c|c|c|c|}
\hline
survey          &early-type &  late(S) &  AGN & others & $N_{tot}$ \\
                & \%       & \%  & \% & \% & \\
\hline
$R < 18.5$       &           &          &      &               &          \\
B93             & $4 \pm 3$  &  $64 \pm 12$  & $13 \pm 5$ & $18 \pm 6$  & 45 \\
PDF             & $20 \pm 5$ & $ 54 \pm 9$ & $10 \pm 4$  & $17 \pm 5$ &  71 \\
\hline
$R > 18.5$        &           &           &          &         &     \\
PDF             & $24 \pm 5$ &  $31 \pm 5$ & $15 \pm 4$ & $31 \pm 5$ & 103 \\
MF              &  $47 \pm 16$ & $37 \pm 14$ & $16 \pm 9$ &  --      & 19 \\
\hline
\end{tabular}
\end{center}
\end{table}

In Table~2 we summarize the composition of the mJy population, according to
the data presented in the most recent works (see Tab.~1 for references).
The survey name is in the first column; 
the fraction (\%) of objects in each class is given in the following columns.
To make a correct comparison of the different works we selected homogeneous 
sub--samples ($S_{1.4GHz}~>$ 1 mJy and magnitude $R~<~18.5$).
The last column of Tab.~2 lists the number of objects belonging 
the each sub--sample. 

The comparison of the data presented in Tab.~2 shows that about 50\% of the 
mJy population is represented by early-type galaxies. For the other 
components the values are not in complete agreement, even if we must underline 
that the differences are not statistically significant, due to the large 
errors.
It is noteworthy that the high number (24\%) of objects which could not be
classified in the ATESP-EIS is paradoxically due to the high quality of the 
spectra. Most of them are peculiar objects with very complex spectral 
features, which makes difficult their classification. However the analysis of 
the radio and optical properties of these objects has shown that probably 
this class is equally composed by early-type galaxies, late(S) galaxies 
and AGN.

Table~3 lists the same quantities as in Tab.~2 for the sub-mJy sources. Here 
we distinguish between two optical magnitude bins ($R<18.5$ and $R>18.5$).
The $R<18.5$ bin in Tab.~3 can be directly compared with the statistics 
reported in Tab.~2. It
is evident that there is a change in the dominant class going from mJy to 
sub--mJy sources: now star forming galaxies constitute half of the population.
This result is in agreement with the hypothesis that the mJy population
is the faint tail of the population (ellipticals and AGN) which dominate at 
high fluxes; while in the sub-mJy region a new population emerges 
where the radio emission is due to star formation phenomenon.

The $R>18.5$ bin in Tab.~3 shows how the sub--mJy population changes 
going from bright ($R<18.5$) to fainter magnitudes. 
While the fraction (10-15\%) of AGN is 
almost independent on the optical magnitude, the contribution of star forming 
galaxies decreases in this faint optical bin from $\sim 1/2$ to $\sim 1/3$ of 
the whole 
population. However there is not complete agreement on the dominant class:
the MF sample is dominated by early-type galaxies, whereas in the PDF 
sample (Georgakakis et al.1999) the importance of star--forming and 
early--type 
galaxies is very similar. Nevertheless, it is worth to note that the fraction 
of $R>18.5$ objects which are not classified in the PDF becomes very large  
(31\%), making very difficult to draw any firm conclusion about the relative 
importance of the different classes in this sample. To summarize, it seems 
that: 

{\it a)} radio sources with $S<1$ mJy and {\it bright optical counterpart} 
are mainly {\it star forming galaxies}; 

{\it b)} radio sources with $S<1$ mJy and {\it faint optical counterpart} are 
mainly {\it early-type galaxies}.

\vspace{0.3cm}
Discrepancies are found also at fainter fluxes. 
The population in the microJy range was investigated by Hammer et al. (1995)
and Richards et al. (1998). Again the samples are complete but very small 
(36 and 29 objects respectively) and the statistics very poor.

Hammer et al., using redshifts color and radio spectral indices, found that 
40\% of the sources probably lie at $z~>$ 1, and nearly half are early-type 
galaxies, frequently containing a low-power AGN nuclear source. 
They concluded that, since
the space density of AGN-driven sources apparently overtakes those powered 
by stellar emission, the contribution of AGN light to the faint source counts
should be re-evaluated.

On the contrary, Richards et al. concluded that the microJy radio galaxies are
distributed over a wide range of redshifts (0.1$<z<$3) and are primarily 
composed of spiral and irregular/merging systems (70-90\%) at modest redshifts
(0$<z<$1). They concluded that the microJy sources are mainly located in 
nearby normal host galaxies, whose bolometric luminosity is dominated by 
starlight rather than an AGN.

To conclude, it is clear that more work is needed in order to understand
nature and distance distribution of the faint radio population. 
In particular it is important to get {\it complete} identification  
and (good quality) spectroscopy follow--up for a deep radio sample, 
large enough to allow a reliable statistical study. 

\vspace{1cm}
{\it Aknowledgements} The authors thank their collaborators: P. Parma, 
H. de Ruiter, M. Wieringa, R. Ekers and especially G. Vettolani, who read 
and commented an earlier version of this manuscript.

\end{document}